\documentstyle[nips97e,times]{article}
\include{epsf}
\title{Synaptic Transmission: An Information-Theoretic Perspective}
\begin{document}
\author{Amit Manwani and Christof Koch \\
Computation and Neural Systems Program\\ 
California Institute of Technology\\
Pasadena, CA 91125 \\
email: quixote@klab.caltech.edu \\
       koch@klab.caltech.edu
}

%

\maketitle
\begin{abstract}
Here we analyze synaptic   transmission from an  information-theoretic
perspective. We derive closed-form expressions for the lower-bounds on
the capacity   of a  simple model  of  a  cortical  synapse under  two
explicit coding  paradigms. Under  the ``signal estimation'' paradigm,
we   assume the signal   to be encoded in  the  mean  firing rate of a
Poisson neuron. The performance of  an optimal linear estimator of the
signal  then provides  a  lower   bound on  the capacity  for   signal
estimation.  Under the ``signal  detection'' paradigm, the presence or
absence  of the signal has to  be detected. Performance of the optimal
spike detector allows us to compute a  lower bound on the capacity for
signal detection.   We find   that single synapses   (for  empirically
measured parameter values) transmit information poorly but significant
improvement can be achieved with a small amount of redundancy.
\end{abstract}

\def \eps0 {\epsilon_{0}}
\def \eps1 {\epsilon_{1}}
\def \w {\omega}
\def \l {\lambda}

\section{Introduction}

Tools   from estimation and  information   theory  have recently  been
applied by  researchers  (Bialek {\em et.  al},  1991) to quantify how
well neurons transmit information  about their random inputs  in their
spike  outputs. In  these approaches,   the neuron is   treated like a
black-box, characterized   empirically   by a   set  of   input-output
records. This ignores the  specific  nature of neuronal processing  in
terms of its known biophysical properties. However, a systematic study
of processing at various stages in a biophysically faithful model of a
single neuron  should be able  to identify  the role  of each stage in
information   transfer in terms of   the  parameters  relating to  the
neuron's dendritic   structure, its  spiking   mechanism, {\em   etc}.
Employing  this reductionist   approach,   we  focus on   a  important
component  of neural processing,  the  synapse, and  analyze a  simple
model  of  a cortical   synapse  under two  different representational
paradigms. Under the  ``signal  estimation'' paradigm, we  assume that
the input  signal  is linearly encoded  in  the mean firing rate  of a
Poisson neuron and the mean-square  error in the reconstruction of the
signal   from    the     post-synaptic   voltage quantifies     system
performance.  From the performance of the  optimal linear estimator of
the signal, a lower bound on the capacity for signal estimation can be
computed.  Under  the  ``signal detection''  paradigm,  we assume that
information is  encoded in  an all-or-none  format  and the  error  in
deciding whether or not a  presynaptic spike occurred by observing the
post-synaptic voltage  quantifies system performance.  This is similar
to the  conventional absent/present(Yes-No) decision paradigm  used in
psychophysics. Performance of the optimal spike  detector in this case
allows us to   compute a  lower  bound on   the capacity  for   signal
detection.

\begin{figure}[h]
\centerline{\epsfxsize=5in \epsfbox{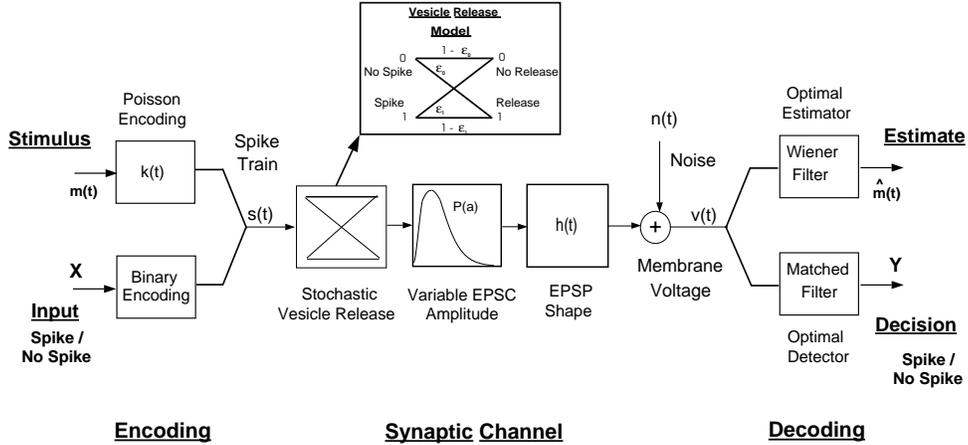}} {\small
\caption{Schematic  block   diagram for    the signal    detection and
estimation tasks. The synapse is modeled  as a binary channel followed
by a filter  $h(t) =  a  \ t \ exp(-t/t_s)$.  where  $a$ is a   random
variable with probability density,  $P(a) = \alpha \ (\alpha  a)^{k-1}
exp(-\alpha a)/ (k-1)!$. The  binary  channel, (inset, $\epsilon_0$  =
Pr[spontaneous release], $\epsilon_1$  = Pr [release failure])  models
probabilistic vesicle release and $h(t)$ models the variable epsp size
observed for  cortical synapses. $n(t)$ denotes additive post-synaptic
voltage noise and is assumed to be Gaussian and white over a bandwidth
$B_n$.  Performance of  the  optimal  linear  estimator  ({\em  Wiener
Filter}) and  the    optimal spike  detector  ({\em Matched   Filter})
quantify synaptic  efficacy   for  signal   estimation and   detection
respectively.}}
\end{figure}

\section{The Synaptic Channel}

Synaptic transmission in cortical neurons is known to be highly random
though the role of  this variability in  neural computation and coding
is still  unclear. In central synapses,  each synaptic bouton contains
only a single  active  release zone,  as opposed  to  the hundreds  or
thousands  found     at   the   much more   reliable     neuromuscular
junction. Thus, in response to an action  potential in the presynaptic
terminal   at most  one    vesicle  is   released (Korn   and   Faber,
1991). Moreover, the probability of vesicle release $p$ is known to be
generally  low (0.1 to   0.4)  from {\em  in  vitro}  studies  in some
vertebrate and   invertebrate   systems    (Stevens,    1994).    This
unreliability is further  compounded by the trial-to-trial variability
in the amplitude of the post-synaptic  response to a vesicular release
(Bekkers {\em et. al}, 1990). In some  cases, the variance in the size
of EPSP is as large as the mean. The empirically measured distribution
of amplitudes is usually skewed to the  right (possibly biased due the
inability of  measuring  very small events)  and can  be  modeled by a
Gamma distribution.

In light  of the   above, we  model  the synapse  as a  binary channel
cascaded by  a random amplitude  filter  (Fig.~1). The binary  channel
accounts for  the  probabilistic vesicle release.  $\epsilon_{0}$  and
$\epsilon_{1}$ denote the probabilities of spontaneous vesicle release
and failure respectively. We follow the binary channel convention used
in digital  communications  ($\eps1  =  1-p$), whereas,  $p$  is  more
commonly  used   in  neurobiology. The  filter   $h(t)$  is chosen  to
correspond  to  the  epsp profile of   a fast  AMPA-like synapse.  The
amplitude of the filter $a$ is modeled as random variable with density
$P(a)$, mean $\mu_{a}$    and standard deviation  $\sigma_a$.   The CV
(standard deviation/mean) of  the  distribution is denoted by  $CV_a$.
We also assume   that additive Gaussian  voltage  noise $n(t)$  at the
post-synaptic   site further corrupts   the epsp  response. $n(t)$  is
assumed to white with  variance $\sigma_{n}^2$ and a bandwidth $B_{n}$
corresponding to the membrane time  constant $\tau$. One can define an
effective  signal-to-noise ratio, $SNR   = E_{a}/N_{o}$, given  by the
ratio  of the  energy in  the epsp  pulse, $E_{h}  = \int_{0}^{\infty}
h^{2}(t) \ dt$     to the noise   power  spectral  density,  $N_{o}  =
\sigma_{n}^{2}/B_{n}$. The performance of  the synapse depends on  the
$SNR$ and  not on the absolute values  of  $E_{h}$ or $\sigma_{n}$. In
the  above model, by   regarding synaptic parameters  as constants, we
have   tacitly ignored  history  dependent  effects  like paired-pulse
facilitation, vesicle  depletion, calcium buffering,  etc, which endow
the  synapse  with  the nature   of  a sophisticated nonlinear  filter
(Markram and Tsodyks, 1997).

\begin{figure}[h]
\begin{tabular}{ll}
\parbox{3in}{\epsfxsize=3in \epsfbox{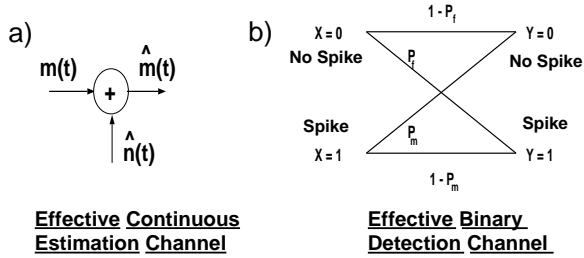}} &
\parbox{1.75in}{{\small \caption{  (a)  Effective  channel  model  for
signal   estimation.   $m(t)$, $\hat{m}(t)$,  $\hat{n}(t)$  denote the
stimulus, the best   linear  estimate, and  the  reconstruction  noise
respectively.  (b) Effective  channel model for  signal detection. $X$
and $Y$ denote the binary variables corresponding to the input and the
decision   respectively.  $P_f$ and  $P_m$    are the effective  error
probabilities.}}}
\end{tabular}
\end{figure}

\section{Signal Estimation}

Let us  assume that the spike  train of the  presynaptic neuron can be
modeled as a doubly stochastic Poisson process with a rate $\lambda(t)
= k(t)*m(t)$ given as a convolution between the  stimulus $m(t)$ and a
filter $k(t)$.  The stimulus is  drawn from a probability distribution
which we assume to  be Gaussian.  $k(t)  = exp(-t/\tau)$ is a low-pass
filter   which  models  the  phenomenological  relationship  between a
neuron's  firing rate  and  its  input current.  $\tau$  is  chosen to
correspond to the membrane time constant. The exact  form of $k(t)$ is
not crucial   and the above  form is  assumed primarily for analytical
tractability.  The   objective is  to  find  the optimal  estimator of
$m(t)$ from the post-synaptic voltage $v(t)$, where optimality is in a
least-mean square sense.  The  optimal  mean-square estimator is,   in
general, non-linear  and reduces to a linear  filter only when all the
signals   and noises are Gaussian.  However,   instead of making  this
assumption, we   restrict ourselves  to the   analysis of  the optimal
linear  estimator, $\hat{m}(t)  =  g(t)*v(t)$, {\em  i.e.}  the filter
$g(t)$ which   minimizes the mean-square  error $E  = \langle  (m(t) -
\hat{m}(t))^{2}\rangle$ where  $\langle . \rangle$ denotes an ensemble
average.   The  overall estimation  system  shown   in  Fig.~1  can be
characterized   by an  effective  continuous  channel (Fig.~2a)  where
$\hat{n}(t) = \hat{m}(t) - m(t)$  denotes the effective reconstruction
noise. System  performance can be quantified  by  E, the  lower E, the
better  the synapse  at  signal transmission.  The expression for  the
optimal  filter  ({\em Wiener filter})  in  the frequency  domain is $
g(\omega) =  S_{mv}(-\omega)/S_{vv}(\omega)$ where $S_{mv}(\omega)$ is
the cross-spectral density (Fourier transform of the cross-correlation
$ R_{mv}$) of $m(t)$   and $s(t)$ and   $S_{vv}(\omega)$ is the  power
spectral density of $v(t)$. The minimum mean-square  error is given by
,    $E      =   \sigma_{m}^{2}       -      \int_{\cal    S}     \mid
S_{mv}(\omega)\mid^{2}/S_{vv}(\omega)\ d \omega $.  The set $ {\cal S}
= \{ \omega \mid S_{vv}(\omega) \neq 0\}$  is called the {\em support}
of $S_{vv}(\omega)$.

Another measure  of system performance  is the mutual information rate
$I(m;v)$ between $m(t)$ and $v(t)$, defined as the rate of information
transmitted by $v(t)$ about $s(t)$.  By the Data Processing inequality
(Cover  1991),   $I(m,v)   \geq   I(m,\hat{m})$. A  lower     bound of
$I(m,\hat{m})$ and thus of $I(m;v)$  is given by the simple expression
$I_{lb}             =             \frac{1}{2}               \int_{\cal
S}\log_{2}[\frac{S_{mm}(\omega)}{S_{\hat{n}   \hat{n}}(\omega)}]     \
d\omega $   (units of bits/sec).    The lower  bound is  achieved when
$\hat{n}(t)$  is  Gaussian and  is independent of   $m(t)$.  Since the
spike train $s(t)  = \sum \delta(t-t_{i})$  is a Poisson process  with
rate  $k(t)*m(t)$,   its power spectrum is  given   by the expression,
$S_{ss}(\omega)=\bar{\lambda} + \mid K(\omega)\mid^{2} S_{mm}(\omega)$
where $\bar{\lambda}$ is the mean firing rate. We assume that the mean
($\mu_{m}$) and variance ($\sigma^{2}_{m}$)  of $m(t)$ are chosen such
that the probability that $\lambda(t) < 0 $ is negligible\footnote{ We
choose   $\mu_m$  and   $\sigma_m$   so  that   $\bar{\lambda}    =  3
\sigma_{\lambda}$ (std of $\lambda$)  so that Prob[$\lambda(t) \leq 0]
< 0.01$.} The vesicle release process is the  spike train gated by the
binary channel   and  so it  is   also  a  Poisson process   with rate
$(1-\epsilon_{1} ) \lambda(t)$. Since $v(t) =  \sum a_{i} h(t - t_{i})
+ n(t)$ is  a filtered Poisson process, its  power spectral density is
given  by  $S_{vv}(\omega)   =  $\mbox{$\mid  H(\omega)\mid^{2}$}  $\{
(\mu_{a}^{2}  + \sigma_{a}^{2})   (1 -  \epsilon_{1})  \bar{\lambda} +
\mu_{a}^{2} (1-\epsilon_{1})^{2}$ \mbox{$\mid  K(\omega)    \mid^{2}$}
$S_{mm}(\omega)  \}  + S_{nn}(\omega)$. The cross-spectral  density is
given by the   expression $S_{vm}(\omega)  = (1-\epsilon_{1})  \mu_{a}
S_{mm}(\omega)  H(\omega)  K(\omega)$.  This allows us   to  write the
mean-square error as,

\[ 
E = \sigma_{m}^{2} - \int_{\cal S} \frac{S_{mm} ^{2}(\omega)}
{\lambda_{eff}(\omega) + S_{mm}(\omega) + S_{eff}(\omega)} \ d\omega 
\]
\[ 
\lambda_{eff}(\omega) = \frac{\bar{\lambda}(1 + CV_{a}^{2})}
{(1-\epsilon_{1}) \ \mid K(\omega) \mid^{2}} \ ,\ S_{eff}(\omega) = 
\frac{S_{nn}(\omega)}{(1-\epsilon_{1})^{2} \mu_{a}^{2} \mid 
H(\omega) \mid^{2} \, \mid K(\omega) \mid^{2}}
\]

Thus,   the  power  spectral density   of   $\hat{n}(t)$ is  given  by
$S_{\hat{n}  \hat{n}} = \l_{eff}(\w)  + S_{eff}(\w)$. Notice that if $
K(\w) \rightarrow \infty,  E \rightarrow   0  $ {\em i.e.}     perfect
reconstruction takes place in the limit of high  firing rates. For the
parameter  values chosen, $S_{eff}(\w)  \ll \l_{eff}(\w)$,  and can be
ignored. Consequently, signal estimation  is {\em shot  noise} limited
and synaptic variability increases shot  noise by a factor $N_{syn}  =
(1+ CV_{a}^2)/(1 - \epsilon_1)$. For $CV_{a} =  0.6$ and $\eps1 = 0.6,
N_{syn}  = 3.4$,  and for  $CV_{a} = 1$  and  $\eps1 =  0.6, N_{syn} =
5$. If  $m(t)$  is chosen  to be  white,  band-limited to  $B_{m}$ Hz,
closed-form  expressions  for $E$ and $I_{lb}$  can   be obtained. The
expression for $I_{lb}$ is tedious and provides  little insight and so
we present only the expression for $E$ below.

\[ 
E(\gamma,B_{T}) = \sigma^{2}_{m} [ 1 - \frac{\gamma}{\sqrt{1 + \gamma}} 
\frac{1}{B_{T}} tan^{-1}(\frac{B_{T}}{\sqrt{1 + \gamma}})] \] 
\[ \gamma = \frac{\sigma_{m}^2 \bar{\l}}{2 \mu_{m}^2 N_{syn} B_{m}}\ ,
\ B_{T} = 2 \pi B_{m} \tau \]

$E$  is a  monotonic   function of $\gamma$  (decreasing)  and $B_{T}$
(increasing).   $\gamma$ can be  considered as the effective number of
spikes available per unit signal bandwidth and $B_{T}$ is the ratio of
the signal  bandwidth and the  neuron  bandwidth. Plots of  normalized
reconstruction error $E_r = E/\sigma^{2}_{m}$ and $I_{lb}$ versus mean
firing   rate ($\bar{\l}$) for  different  values  of signal bandwidth
$B_{m}$ are shown in Fig.~3a  and  Fig.~3b respectively. Observe  that
$I_{lb}$  (bits/sec) is insensitive  to $B_{m}$ for  firing rates upto
200Hz because the decrease  in  quality of estimation  ($E$  increases
with $B_{m}$)  is  compensated by   an   increase in  the  number   of
independent samples ($2B_{m}$)  available per second.  This phenomenon
is characteristic of systems operating in the low SNR regime. $I_{lb}$
has the generic form, $I_{lb} = B \ \log(1 + S/(N B))$, where $B$, $S$
and  $N$  denote  signal   bandwidth,  signal  power  and  noise power
respectively. For  low  SNR,  $I  \approx B \    S/(N B) =   S/N$,  is
independent of B. So one can argue that, for our choice of parameters,
a single synapse is a  low SNR system.   The analysis generalizes very
easily to the  case of multiple  synapses where all  are driven by the
same  signal  $s(t)$. (Manwani and  Koch,   in preparation).  However,
instead  of    presenting the rigorous analysis,    we   appeal to the
intuition gained from the single synapse case.  Since a single synapse
can be regarded as  a shot noise source, $n$  parallel synapses can be
treated as   $n$ parallel noise  sources.  Let us  make  the plausible
assumption that these noises are  uncorrelated. If optimal  estimation
is  carried out  separately  for each synapse   and the estimates  are
combined optimally,  the  effective  noise variance   is  given by the
harmonic  mean  of  the     individual   variances {\em   i.e.}    $1/
\sigma^{2}_{neff}     =  \sum_{i}1/\sigma^{2}_{ni}$.  However,  if the
noises  are added first and   optimal  estimation is carried out  with
respect  to the  sum,  the effective  noise variance is   given by the
arithmetic   mean    of   the     individual  variances,  {\em   i.e.}
$\sigma^{2}_{neff}  = \sum_{i} \sigma^{2}_{ni}/ n^{2}  $. If we assume
that all synapses are  similar   so that $\sigma_{ni}^2 =   \sigma^2$,
$\sigma^{2}_{neff}  = \sigma^{2}/n$. Plots  of  $E_r$ and $I_{lb}$ for
the  case of 5  identical  synapses are shown  in  Fig.~3c and Fig.~3d
respectively. Notice that  $I_{lb}$ increases with  $B_{m}$ suggesting
that the system  is no longer in the   low SNR regime.  Thus, though a
single synapse has   very low capacity,  a small  amount of redundancy
causes a considerable increase in performance. This is consistent with
the  fact  the in  the low $SNR$   regime, $I$ increases linearly with
$SNR$, consequently, linearly with $n$, the number of synapses.

\begin{figure}[h]
\centerline{\epsfxsize=5in \epsfbox{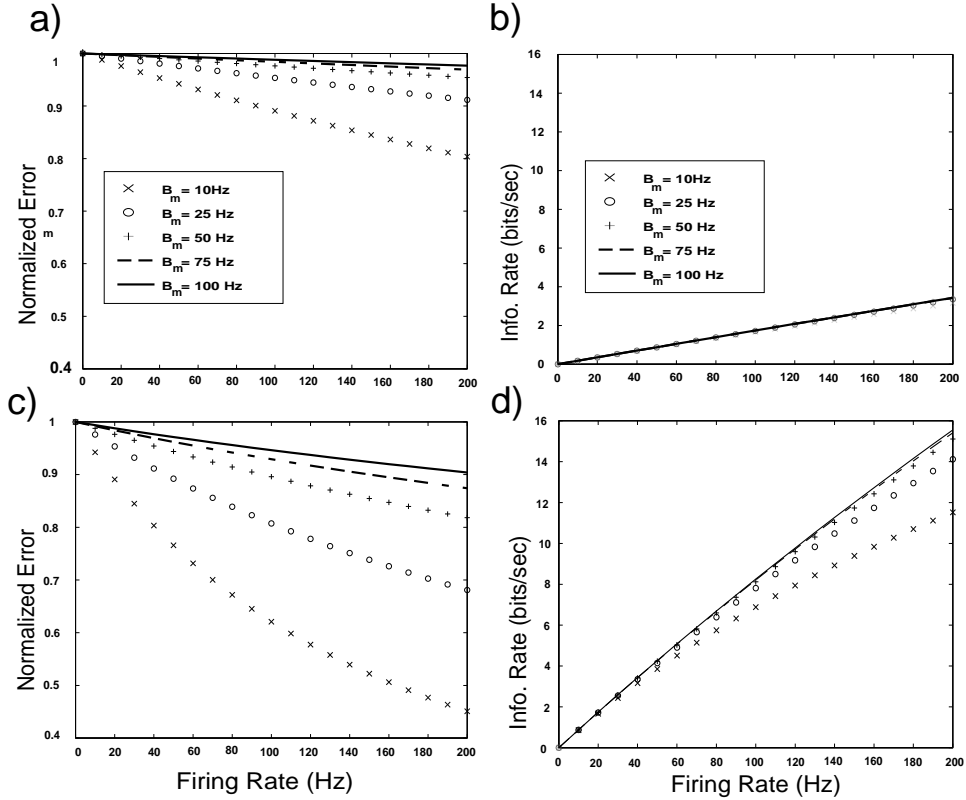}}
{\caption{\small $E_{r}$ and $I_{lb}$ vs. mean firing rate
($\bar{\lambda}$) for n = 1 [(a) and (b)] and n = 5 [(c) and (d)]
identical synapses respectively (different values of $B_m$) for signal 
estimation. Parameter values are 
$\epsilon_{1} =0.6$, $\epsilon_0 = 0$, $CV_a$ = 0.6, $t_s$ = 0.5 msec, 
$\tau$ = 10msec, $\sigma_n$ = 0.1 mV, $B_n$ = 100 Hz.}}
\end{figure}

\section{Signal Detection}

The goal in signal detection is  to decide which  member from a finite
set of signals was generated by a source, on the basis of measurements
related  to  the  output only in    a statistical  sense.  Our example
corresponds to its simplest case, that of {\em binary detection}.  The
objective is  to derive   an   optimal spike   detector based on   the
post-synaptic  voltage in  a  given time  interval.  The criterion  of
optimality  is minimum probability of  error ($P_{e}$).  A false alarm
(FA)  error  occurs  when a spike   is falsely  detected even  when no
presynaptic  spike occurs and  a  miss error (M)   occurs when a spike
fails to be detected.  The probabilities of  the errors are denoted by
$P_{f}$ and $P_{m}$ respectively. Thus,  $P_{e} = (1-p_o)\ P_{f} + p_o
\ P_{m}$  where $p_o$  denotes the  a  priori  probability of a  spike
occurrence.   Let   $X$ and $Y$  be  binary   variables denoting spike
occurrence   and  the decision respectively.  Thus,   $X=1$ if a spike
occurred else $X=0$. Similarly, $Y =  1$ expresses the decision that a
spike occurred.   The {\em posterior  likelihood} ratio  is defined as
${\cal L}(v) = Pr(v \mid X=1)/Pr(v \mid X=0)$ and the prior likelihood
as ${\cal L}_{o} = (1 - p_o)/p_o$.  The optimal spike detector employs
the well-known {\em  likelihood ratio } test  , ``If ${\cal L}(v) \geq
{\cal L}_{o}$ {\bf Y = 1} else {\bf  Y = 0}''.  When $X=1$, $v(t)= a \
h(t) +  n(t)$ else $v(t)  = n(t)$.  Since  $a$ is  a random  variable,
${\cal L}(v) = (\int Pr(v\mid X=1;a) \ P(a)\ da )/ Pr(v\mid X=0)$.  If
the noise $n(t)$   is Gaussian and white,  it  can be  shown  that the
optimal decision rule  reduces to a {\em  matched filter}\footnote{For
deterministic  $a$, the result   is well-known, but  even if  $a$ is a
one-sided random  variable,  the matched  filter can   be  shown to be
optimal.}, {\em  i.e.}   if the correlation,  $r$   between $v(t)$ and
$h(t)$ exceeds  a particular threshold  (denoted by  $\eta$),  $Y = 1$
else $Y= 0$.  The overall   decision  system shown  in  Fig.~1 can  be
treated as effective binary  channel (Fig.~2b). The system performance
can  be  quantified   either  by  $P_{e}$   or $I(X;Y)$,   the  mutual
information between   the binary random  variables,  $X$ and $Y$. Note
that even when $n(t) =  0 \ (SNR =\infty)$,  $P_{e} \neq 0$ due to the
unreliability  of   vesicular release.    Let $P^{*}_{e}$   denote the
probability  of  error when $SNR  =\infty$.    If $\epsilon_{0} =  0$,
$P^{*}_{e}  =p_o \  \epsilon_{1}$ is  the  minimum possible  detection
error. Let $P^{o}_{f}$ and $P^{o}_{m}$ denote FA and M errors when the
release is ideal ($\epsilon_{1}=0,\   \epsilon_{0}  = 0$). It can   be
shown that

\[ 
P_{e} = P^{*}_{e} + P^{o}_{m}[p_o(1-\epsilon_{1}) - (1-p_o)
\epsilon_{0}] + P^{o}_{f}[(1-p_o)(1-\epsilon_{0}) - p_o\epsilon_{1}] 
\] 
\[ 
P_{f} = P^{o}_{f} \ , \ P_{m} = P^{o}_{m} + \epsilon_{1}(1 - P^{o}_{m} 
+ P^{o}_{f}) 
\]

Both $P^{o}_{f}$ and $P^{o}_{m}$  depend on $\eta$. The  optimal value
of $\eta$   is  chosen such that  $P_{e}$   is minimized.  In general,
$P^{o}_{f}$ and $P^{o}_{m}$ can  not  be expressed in  closed-form and
the optimal $\eta$ is  found  using the  graphical {\em ROC  analysis}
procedure. If we  normalize $a$ such  that $\mu_{a}  = 1$, $P^{o}_{f}$
and  $P^{o}_{m}$   can  be  parametrically expressed  in   terms  of a
normalized threshold $\eta^*$, $P^{o}_{f} = 0.5 [ 1 - Erf(\eta^*)]$, \
$P^{o}_{m} = 0.5 [1 + \int_{0}^{\infty} \ Erf(\eta^* - \sqrt{SNR} \ a)
\  P(a) \ da]$.  $I(X;Y)$  can be computed using  the  formula for the
mutual information for a binary  channel, $I = {\cal H}(p_o\ (1-P_{m})
+ (1-p_o)\  P_{f}) - p_o  {\cal  H}(P_{m}) -  (1-p_o) {\cal H}(P_{f})$
where ${\cal H}(x)  = - x   \log_{2}(x) - (1-x) \log_{2}(1-x)$ is  the
binary entropy function.  The analysis can  be generalized to the case
of $n$ synapses  but the expressions involve $n$-dimensional integrals
which need to be evaluated numerically. The  Central Limit Theorem can
be used to simplify  the case of very  large $n$. Plots of $P_{e}$ and
$I(X;Y)$ versus $n$ for different values  of $SNR$ (1,10,$\infty$) for
the case   of identical  synapses are  shown   in Fig.~4a and  Fig.~4b
respectively.  Yet again, we observe the poor  performance of a single
synapse and the substantial improvement due  to redundancy. The linear
increase of $I$ with $n$ is similar to the  result obtained for signal
estimation.

\begin{figure}[h]
\centerline{\epsfxsize=5in \epsfbox{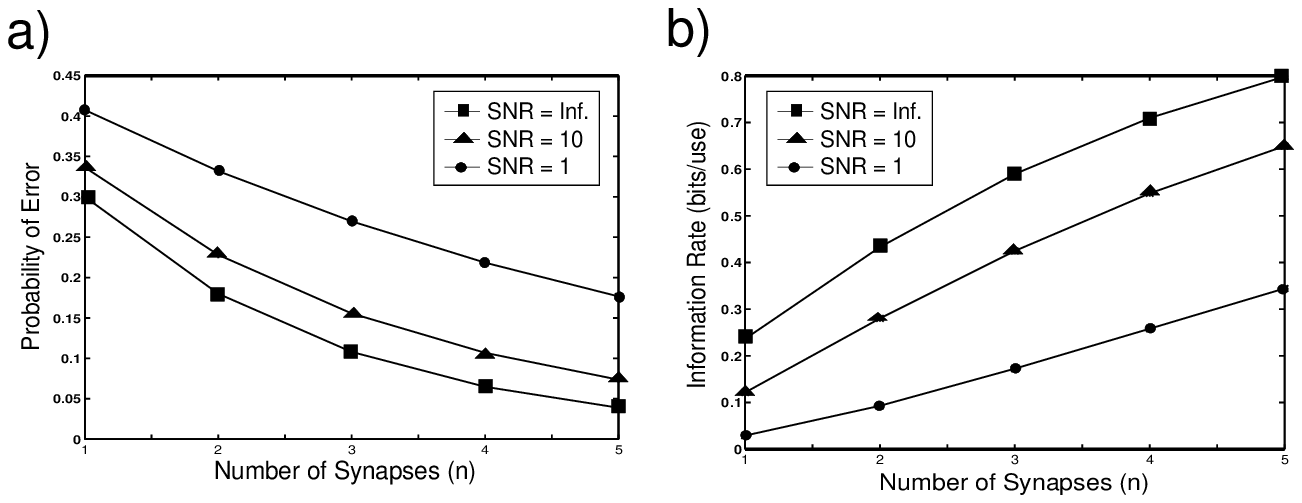}} {\small{$P_{e}$  (a)
and $I_{lb}$ (b) vs. the number of synapses, $n$, (different values of
$SNR$)   for signal  detection.   $SNR$  =   Inf. corresponds  to   no
post-synaptic voltage   noise.  All the synapses   are   assumed to be
identical.  Parameter values are $p_o$ =  0.5, $\epsilon_{1}  = 0.6 $,
$\epsilon_0  = 0$, $CV_a$  = 0.6, $t_s$ = 0.5  msec, $\tau$ = 10 msec,
$\sigma_n$ = 0.1 mV, $B_n$ = 100 Hz.}}
\end{figure}

\section{Conclusions}

We find that a single synapse is rather ineffective as a communication
device but with a little redundancy neuronal communication can be made
much more robust. Infact, a single synapse can  be considered as a low
SNR device, while  5 independent synapses in  parallel approach a high
SNR  system.  This is consistently  echoed  in the  results for signal
estimation and signal detection. The   values of information rates  we
obtain are  very  small  compared   to  numbers  obtained  from   some
peripheral sensory neurons (Rieke {\em et.  al}, 1996).  This could be
due to an over-conservative choice of parameter values  on our part or
could    argue for   the  preponderance   of   redundancy  in   neural
systems. What we have presented above are  preliminary results of work
in progress and so the path ahead is much longer  than the distance we
have covered   so  far. To   the best of   our  knowledge, analysis of
distinct  individual components of a    neuron from an  communications
standpoint has not been carried out before.

\subsubsection*{Acknowledgements}

This research  was supported  by NSF,  NIMH  and the Sloan  Center for
Theoretical Neuroscience.  We thank Fabrizio Gabbiani for illuminating
discussions.

\subsubsection*{References}

Bekkers, J.M., Richerson, G.B.  and  Stevens, C.F. (1990) ``Origin  of
variability  in quantal  size in   cultured  hippocampal neurons   and
hippocampal  slices,'' {\em Proc.  Natl.   Acad. Sci. USA} {\bf   87:}
5359-5362.

Bialek,  W. Rieke, F.  van Steveninck,  R.D.R.  and Warland, D. (1991)
``Reading a neural code,'' {\em Science} {\bf 252:} 1854-1857.

Cover,  T.M.,  and Thomas, J.A.  (1991)   {\em Elements of Information
Theory.} New York: Wiley.

Korn, H.  and  Faber, D.S.  (1991)   ``Quantal analysis  and  synaptic
efficacy in the CNS,'' {\em Trends Neurosci.} {\bf 14:} 439-445.

Markram,  H.  and  Tsodyks,  T.   (1996)  ``Redistibution of  synaptic
efficacy between neocortical  pyramidal  neurons,'' {\em  Nature} {\bf
382:} 807-810.

Rieke, F.   Warland, D. van  Steveninck, R.D.R.  and Bialek, W. (1996)
{\em Spikes: Exploring the Neural Code.} Cambridge: MIT Press.

Stevens, C.F. (1994) ``What form should  a cortical theory take,'' In:
{\em Large-Scale Neuronal Theories of the Brain,}  Koch, C. and Davis,
J.L., eds., pp. 239-256.  Cambridge: MIT Press.

\end{document}